\documentclass[%
reprint,
10pt,
nofootinbib,
amsmath,amssymb,
aps,
]{revtex4-1}

\usepackage{graphicx}
\usepackage{dcolumn}
\usepackage{bm}
\usepackage{footmisc}
\usepackage{amsmath,amssymb} 
\usepackage{mathrsfs}
\usepackage{amsfonts}
\usepackage{pgfplots}
\pgfplotsset{compat=1.5}
\usepackage{floatrow}
\usepackage{soul}
\usepackage{xcolor}

\def\IB#1{\boldsymbol{#1}} 
\def\W/!i#1{\Wi} 
\def\bten#1{\IB{\mathsf{#1}}}

\begin{document}
	
	\preprint{APS/123-QED}
	
	\title{On the cross-streamline lift of microswimmers in viscoelastic flows}
	
	\author{Akash Choudhary\textsuperscript{1}}
	\homepage{a.choudhary@campus.tu-berlin.de}
	\author{Holger Stark\textsuperscript{1}}
	\homepage{holger.stark@tu-berlin.de}
	\affiliation{%
		\textsuperscript{1}Institute of Theoretical Physics, Technische Universit\"{a}t Berlin, 10623 Berlin, Germany
	}%
	
	%

	
	\begin{abstract}
The current work studies the dynamics of a microswimmer in pressure-driven flow of a weakly viscoelastic fluid. Employing the 
second-order fluid model, we show that the self-propelling swimmer experiences a viscoelastic  swimming lift in addition to  the well-known passive lift that arises from
its resistance to shear flow. Using the reciprocal theorem, we evaluate analytical expressions for the swimming lift experienced by neutral and pusher/puller-type 
swimmers and show that they depend on the hydrodynamic signature associated with the swimming mechanism.
We find that for neutral swimmers focusing towards the centerline is accelerated by two orders of magnitude, while for force-dipole swimmers no net modification in cross-streamline migration occurs.
	\end{abstract}
	
	\maketitle
	


	Biological microswimmers are ubiquitous in polymeric media such as cervical, bronchial, and intestinal mucus films \cite{suarez2006sperm,levy2014pulmonary}. 
	During the generation of biofilms, detrimental to bio-engineering and industrial processes, most bacteria release a mixture of proteins, DNA, and polysaccharides, endowing the fluid with viscoelastic properties \cite{persat2015mechanical,conrad2018confined}, which significantly alter the swimmer's dynamics \cite{jabbarzadeh2014swimming}.
	Pathogens like ulcer-causing
	\textit{Helicobacter pylori} thrive by altering the rheology of mucus lining in stomach \cite{celli2009helicobacter}.
	Recent theoretical and experimental studies on the dynamics of motile microorganisms in Newtonian flows have revealed their rich 
	dynamics and ability to swim against fluid flows, which aids in seeking nutrients and  in reproduction \cite{bretherton1961rheotaxis,hill2007hydrodynamic,nash2010run,zottl2012nonlinear,zottl2013periodic,tung2015emergence,Mathijssen19,lauga2020fluid}.
	
	Although recent works have provided insights in self-propulsion in non-Newtonian environments \cite{lauga2007propulsion,shen2011undulatory,zhu2012self,pak2012micropropulsion,keim2012fluid,elfring2015theory,sznitman2015locomotion,li2015undulatory,datt2015squirmingST,li2016collective,li2017near,ives2017mechanism,datt2017activeComplex,zhang2018reduced,zottl2019enhanced,choudhary_2020_Janus,binagia2020swimming,lauga2020fluid}, 
	few have studied the impact of fluid rheology on the dynamics of microswimmers in confined flows \cite{mathijssen2016upstream,corato2017dynamics,ardekani2012emergence}. 
	\citet{mathijssen2016upstream} developed a model to predict the dynamical states of microswimmers in non-Newtonian Poiseuille flows.
	Employing a second-order fluid model, they demonstrated that normal-stress differences reorient the swimmers to cause 
	centerline upstream migration (rheotaxis). 
	This suggests that non-Newtonian properties can help microorganisms evade the boundary accumulation, prevalent in quiescent Newtonian fluids \cite{berke2008hydrodynamic,smith2009human}.

	The evidence of centerline reorientation of microswimmers can be traced back to pioneering works on viscoelastic focusing of passive particles in Poiseuille flows \cite{karnis1966particle,gauthier1971particle,ho1976migration}.
	These studies showed that normal stresses exert a lift force that focuses the particles on the centerline. \citet{ho1976migration} used 
	the reciprocal theorem and derived an analytical expression for this lift in weakly elastic Boger fluids.
	The study suggested that
	the hydrodynamic disturbances around the particle produces a hoop stress
	that, in the presence of non-uniform shear rate, 
	generates a cross-streamline lift.
	Recent progress in electrophoresis \cite{xuan2018,choudhary2020electrokinetically,khair_comment} 
	has also shed light on the importance of hydrodynamic disturbances in determining these lift forces.
	
	Active microswimmers, as opposed to passive particles, generate additional disturbance  flow fields
	in the fluid 
	due to their self-propulsion.
	Therefore, the associated lift force or velocity should also have an active component that is characteristic 
	of the self-propulsion mechanism.
	Since this component is absent 
	in the recently proposed models \cite{mathijssen2016upstream}, in this communication we derive the 
	`swimming lift' in a second-order fluid (SOF), and show 
	how it affects the dynamics of a microswimmer in the Poiseuille flow of 
	a viscoelastic fluid.
	We choose the SOF model because it provides an asymptotic 
	approximation for  a majority of slow and slowly varying viscoelastic flows \cite{leal_1979,bird1987dynamics}. 
	
	Figure\ \ref{fig:schematic} shows  a spherical swimmer of radius $a$ at position $\IB{r}$ that self-propels with  velocity $ \IB{v}_{s} = v_s \IB{p}$ in a two-dimensional 
	pressure-driven flow  $ \IB{v}_{f} = v_{m} [1-({x}/{w})^{2}] \, \IB{e}_{z} $, where $ v_{m} $  is the maximum flow velocity and $ w $ 
	is the half channel width. 
	The flow profile of the second-order fluid is identical to Poiseuille flow but the pressure field varies in $ x- $direction \cite{ho1976migration}.
	In the absence of noise, the swimmer's  dynamics is governed by
	\begin{equation}\label{kin}
		\IB{\dot{r}}  = \IB{p} +  \bar{\IB{v}}_{f}  +\mathcal{F}(x,{\IB{p}}) \, \IB{e}_{x}\, , \quad 
		\IB{\dot{p}}  = \frac{1}{2} \left(\nabla  \times \bar{\IB{v}}_{f}  \right) \times \IB{p}, 
	\end{equation}
	where the velocities are non-dimensionalized by swimming speed $v_s$, lengths by $w$, and time by $ w/v_{s} $.  $ \mathcal{F} $ denotes the total viscoelastic lift velocity, which comprises the passive and swimming lift. 
	Below, we derive the analytical expressions for the lift velocities.
	We note that normal stresses 
	also modify the  particle rotation and the drift velocity along the channel axis.
	We evaluated these modifications and found that they do not play a significant role in determining the swimmer dynamics.

	The inertia-less or creeping flow
	hydrodynamics is governed by the continuity  equations for mass and momentum, which we formulate here
	in the co-moving swimmer frame $ \lbrace \tilde{x},\tilde{y}, \tilde{z}  \rbrace $ as: 
	\begin{align}\label{SOF}
		\tilde{\nabla} \cdot \IB{V} = 0 ,  \;   \tilde{\nabla} \cdot \bten{T}=0 
	\end{align} 
	in order to calculate the viscoelastic lift velocity.
	Length, velocity, and pressure in (\ref{SOF}) 
	are non-dimensionalized by $ a $, $ \kappa v_{m} $, and $ \mu \kappa v_{m}/a $, respectively. 
	Here, $ \kappa $ is the particle to channel width ratio ($ a/2w $) and $ \mu $ is the fluid viscosity.
	In the above equation, $ \bten{T} $ is the total stress tensor 
	of a second-order fluid and thus has the form \cite{bird1987dynamics}:  $  \bten{T} =	-P \, \bten{I} + 2 \, \bten{E} + \text{Wi} \, \bten{S} $.
	Here, $ \bten{E} $ denotes the rate of strain tensor and $ \text{Wi} =  (\Psi_{1} + \Psi_{2})G/\mu $
	is the shear based Weissenberg number,
	where $ \mu $ is the viscosity, 
	$ G = v_m / 2w $ characterizes the shear rate in the
	background flow,
	and 
	$\Psi_1,\Psi_2$ represent the dimensional steady-shear normal stress coefficients that are measured experimentally \cite{bird1987dynamics}.
	The polymeric stress tensor $ \bten{S} =
	4 \bten{E} \cdot \bten{E} + 2 \delta \overset{\Delta}{\bten{E}} $ is non-linear in $\bten{E}$ and contains the lower-convected time derivative of  $\bten{E}$ denoted by
	$ \Delta $.
	The viscometric parameter $ \delta = -\Psi_{1}/2(\Psi_{1} + \Psi_{2}) $ generally varies from $ -0.5 $ to $ -0.7 $ for most viscoelastic fluids \cite{caswell1962creeping,leal1975slow,koch2006stress}.
	
	\begin{figure}
		\centering
		\includegraphics[width=.56\columnwidth]{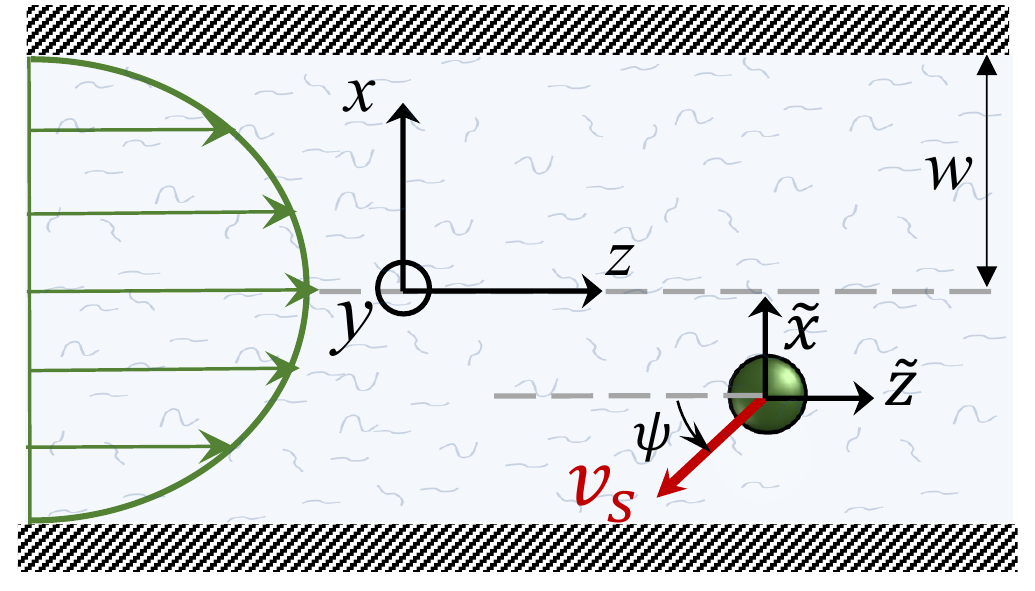}
		\caption{A spherical microswimmer with 	velocity $v_s  \IB{p}$ moves in a pressure-driven flow of a second-order fluid inside a channel with  half width $w$. The coordinate frame $\{ \tilde{x},\tilde{y},\tilde{z} \}$ co-moves	with the swimmer.} 
		\label{fig:schematic}%
	\end{figure}

	
	
	We now focus on determining the lift velocity of a microswimmer that disturbs the background flow in two ways.
	%
	First of all, the microswimmer resists straining by the flow and second, it generates a flow field characteristic of its
	swimming mechanism, 
	for which we first take a source-dipole swimmer.
	We split the full velocity field ($ \IB{V} =\IB{v}^{\infty} + \IB{v}  $) into background flow field $ \IB{v}^{\infty} $ and disturbance field $ \IB{v} $.
	Substituting this in the governing equations (\ref{SOF}) yields:
	\begin{equation}\label{GE_main}
		\tilde{\nabla} \cdot \IB{v} = 0, \quad  - \tilde{\nabla}  p + \tilde{\nabla} ^{2} \IB{v} = - \text{Wi}  (\tilde{\nabla} \cdot \bten{s}) ,
	\end{equation}
	where $ \bten{s} $ is the polymeric stress tensor associated with the disturbance flow field (elaborated in 
	the electronic supplementary material ESI).
	Assuming weak viscoelasticity, we perform a perturbation expansion in Wi and divide Eq.\ (\ref{GE_main}) into two problems: the Stokes equation 
	for the zeroth order of the disturbance field
	and $ - \tilde{\nabla}  p_{1} + \tilde{\nabla} ^{2} \IB{v}_{1} =-  \tilde{\nabla} \cdot \bten{s}_{0} $ at first order.
	Following earlier works on viscoelastic lift \cite{ho1976migration,choudhary2020electrokinetically}, we use the reciprocal theorem to attain the lift velocity from the first-order problem
	\begin{equation} \label{lift}
		\mathcal{F} = - \frac{\text{Wi}}{6\pi} \int_{V_{f}} \bten{s}_{0} : \tilde{\nabla} \IB{u}^{t} \, \text{d}V .
	\end{equation}
	Here, $ \IB{u}^{t} $ is the auxiliary or test velocity field that belongs to a forced particle moving along the $x$-direction with unit velocity in a Newtonian fluid.
	The polymeric stress $ \bten{s}_{0}$  corresponds to the Stokes solution $ \IB{v}_0 $ of the microswimmer consisting of (i) 
	a source-dipole field, which we attain from the squirmer model \cite{Lighthill52,Blake71,zottl2016emergent},
	$ 	\IB{v}_0^{\text{swim}} = \frac{v_{s}\IB{p} }{2 \tilde{r}^{3}} \IB{\cdot} \left[    \frac{3 \tilde{\IB{r}} \tilde{\IB{r}} }{\tilde{r}^{2}}   -\bten{I} \right] $, 
	and (ii) the passive disturbance field $ \IB{v}_0^{\text{passive}} $, which is led by the stresslet; higher order terms are obtained from Lamb's general solution \cite{lamb}.
	Using the corresponding $\bten{s}_0$  in Eq.\ (\ref{lift}), results in the viscoelastic lift velocity given in 
	units of $v_s$: 
	\begin{equation} \label{F_SD}
		\mathcal{F} (x,\psi) =    \text{Wi} \left[\frac{80}{9}x \, \bar{v}_{m} \kappa^{2} (1+3\delta) - x \, (1+ \delta) \cos \psi  \right].
	\end{equation}
	The first component in Eq. (\ref{F_SD}) is the passive lift $ \mathcal{F}_{\text{passive}} $ \cite{ho1976migration}. By fixing $ \delta $ to a widely-used value of $ -0.5 $ (\emph{i.e.} $ \Psi_{2}=0 $), we observe that $ \mathcal{F}_{\text{passive}}$ focuses the particle towards the centerline.
	The second component 
	is the swimming lift $ \mathcal{F}_{swim} $ that arises due to the source-dipole disturbance created by the neutral swimmer. 
	We note two striking features of $ \mathcal{F}_{\text{swim}} $:  the dependence on swimmer orientation through $ \cos \psi $
	and that its magnitude is larger by a factor $\kappa^{-2}$ compared to the first term\footnote{The correction to the drift
		velocity in $ z-$direction is $ -\text{Wi}  \, x \,  \kappa \sin \psi (1+\delta) $,
		and the correction to rotation is found to be $ - \frac{8}{3} \text{Wi} \,\kappa \sin \psi $. We verified that these modifications do not alter the dynamics neither qualitatively nor quantitatively and
		thus neglect their contributions
		for simplicity. Details are provided in the ESI.}.

	\begin{figure}[b]
		\centering
		\includegraphics[width=0.47\textwidth]{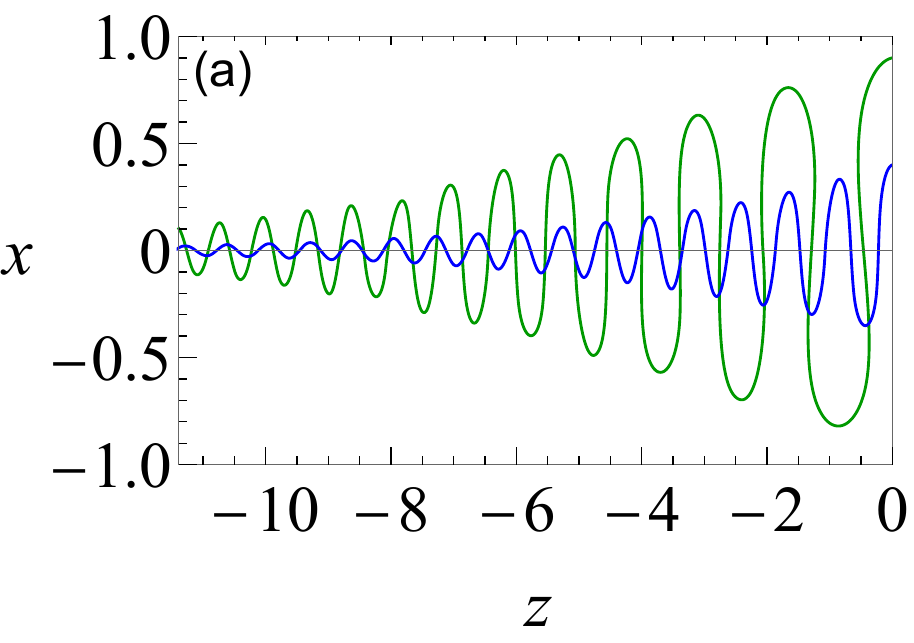} 
		\quad
		\includegraphics[width=0.47\textwidth]{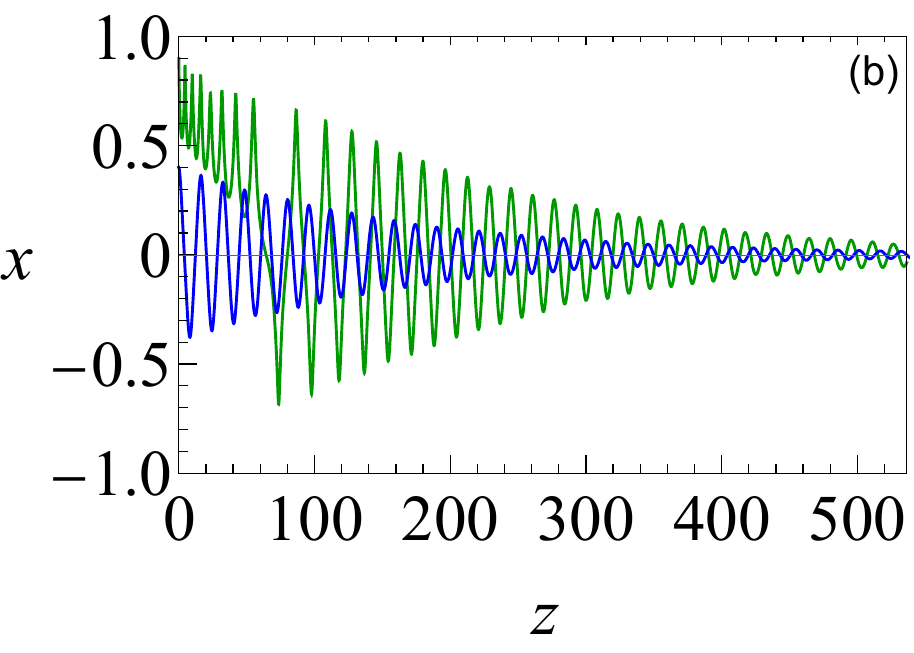}
		\caption{Trajectories showing the centerline focusing of a neutral (source-dipole) 
			swimmer oriented upstream while 		(a) swimming 		upstream ($ \bar{v}_{m}=0.9 $) and (b) 		drifting downstream 
			($\bar{v}_{m} = 8 $). Other parameters: $ \text{Wi}=0.1, \, \kappa=0.1, \, \delta=-0.5. $ The 
			blue and green
			trajectories only differ in the initial condition $ x_{0} $.}%
		\label{fig:Traj_SD}%
	\end{figure}
	
	Now, we substitute $ \mathcal{F}$ 
	into the dynamic equations
	(\ref{kin}) and examine the effect of $ \mathcal{F} _{\text{swim}} $ on the microswimmer dynamics.
	We find two fixed points in the $ x-\psi $ plane at $ x=0 $, with the microswimmer oriented upstream ($ \psi = 0 $) or downstream ($ \psi=\pm \pi $). A linear stability analysis provides the following eigenvalues for these fixed points:
	\begin{align}\label{ev}
		\lambda_{1} \approx& \frac{\text{Wi}}{18} \left[  -9(1+\delta) + 80 (1+3\delta) \kappa^{2} \bar{v}_{m} \right] \pm {\rm i} \, \bar{v}_{m}^{1/2}, \nonumber \\
		\lambda_{2} \approx& \frac{\text{Wi}}{18} \left[  9(1+\delta) + 80 (1+3\delta) \kappa^{2} \bar{v}_{m} \right]  \pm   \, \bar{v}_{m}^{1/2} . 
	\end{align}
	For a typical value of $ \delta=-0.5 $ and weak viscoelasticity limit ($ {\rm Wi} \ll1 $), the downstream swimming corresponds to a saddle fixed point ($ \lambda_{2} $), while the upstream swimming along $ x=0 $ corresponds to a stable fixed point ($ \lambda_{1} $). 
	For $ \delta=-0.5 $, the  sign of the real part of  $\lambda_1$ 
	shows that both swimming and passive lift components stabilize the upstream swimming. However, the strong swimming lift
	can help the neutral microswimmer attain centerline equilibrium more rapidly by a relaxation factor of $ \kappa^{-2} $, as also shown in the trajectories of Fig.\ref{fig:Traj_SD} that are evaluated by substituting (\ref{F_SD}) in (\ref{kin}).

	Now, we shift our attention from neutral squirmers to flagellated microorganisms, such as \textit{E. coli} and \textit{Chlamydomonas}, that generate a force-dipole field at the leading order \cite{pedley1992hydrodynamic,berke2008hydrodynamic}: $ \IB{v}_0 = \mathcal{P}  \IB{r} \big[   \frac{-1}{r^{3}} + 3 \frac{\left(\IB{r} \cdot \IB{p}  \right)^{2} }{r^{5 }}  \big] $.
	Here $ \mathcal{P} $ is the 
	force-dipole strength 
	in units of $ 8\pi \mu a^{2} v_s $, which depends on the swimming mechanism \cite{berke2008hydrodynamic,drescher2010direct,drescher2011fluid}.
	Earlier studies on \textit{E. coli} \cite{drescher2011fluid,chattopadhyay2006swimming} and \textit{Chlamydomonas} \cite{minoura1995strikingly} suggest that $ |\mathcal{P}| $ varies roughly between 0.04 - 0.3.
	Following the procedure outlined for a source-dipole swimmer, we obtain the swimming lift velocity of the force-dipole swimmer as
	\begin{equation} \label{F_FD}
		\mathcal{F}_{\text{swim}}  =     (8/3)  \text{Wi} \, \kappa \, \mathcal{P} (1+3\delta) \sin 2 \psi  ,
	\end{equation}
	and find it to 
	depend on the constant curvature in Poiseuille flow, as detailed in ESI.
	Although this lift is 
	larger 
	than $\mathcal{F}_{\text{passive}}$
	by a factor $\kappa^{-1}$, there is no net lateral motion because the pure angular dependence
	cancels out on average as the particle tumbles continuously.
	The trajectories in Fig.\ref{fig:Traj_FD} 
	(a) and (b)
	show the dynamics of a force-dipole swimmer; the focusing along the centerline 
	is purely due to $ \mathcal{F}_{\text{passive}} $.
	
	\begin{figure}
		\centering
		\includegraphics[width=0.47\textwidth]{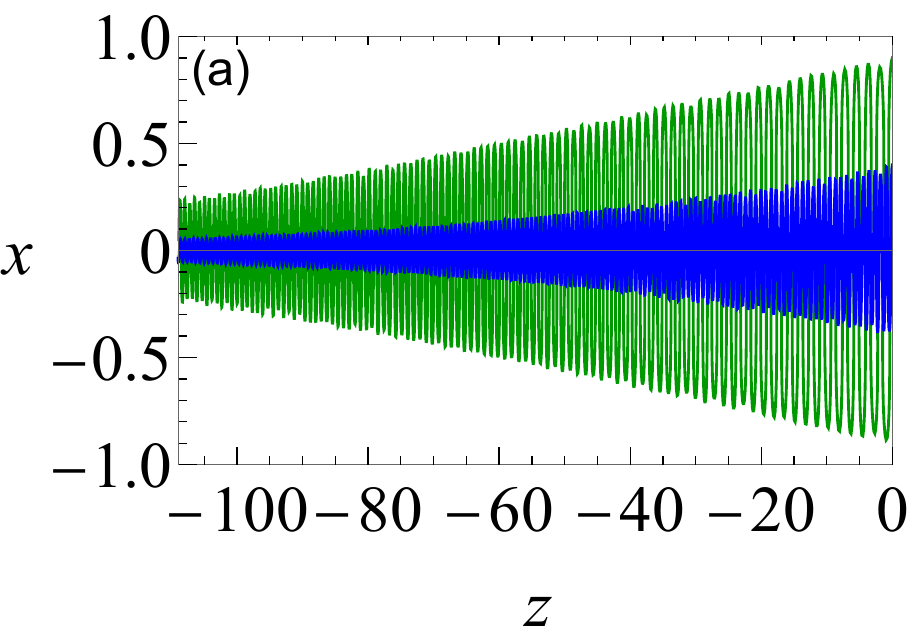} 
		\quad
		\includegraphics[width=0.47\textwidth]{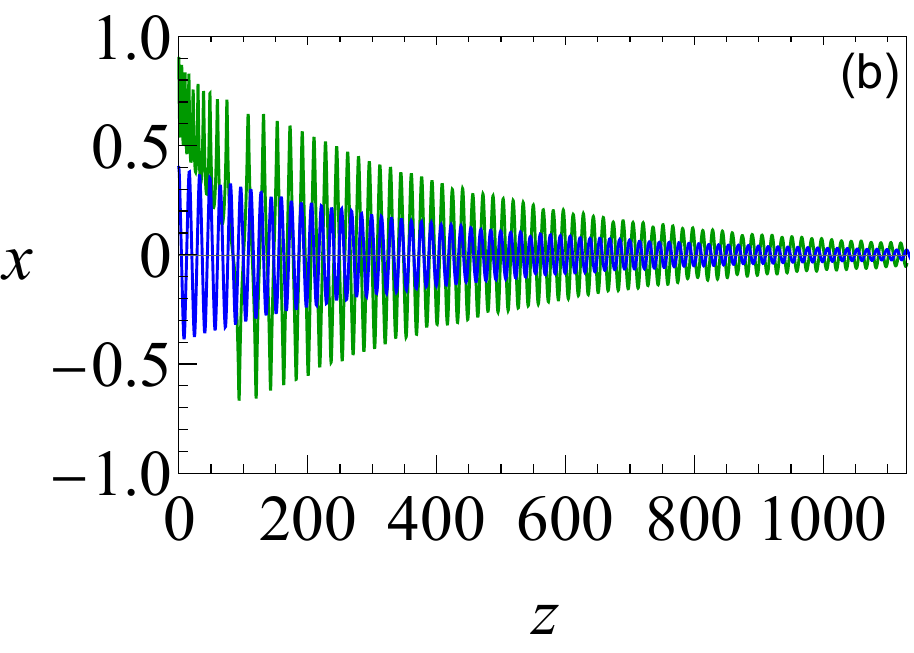}
		\includegraphics[width=0.47\textwidth]{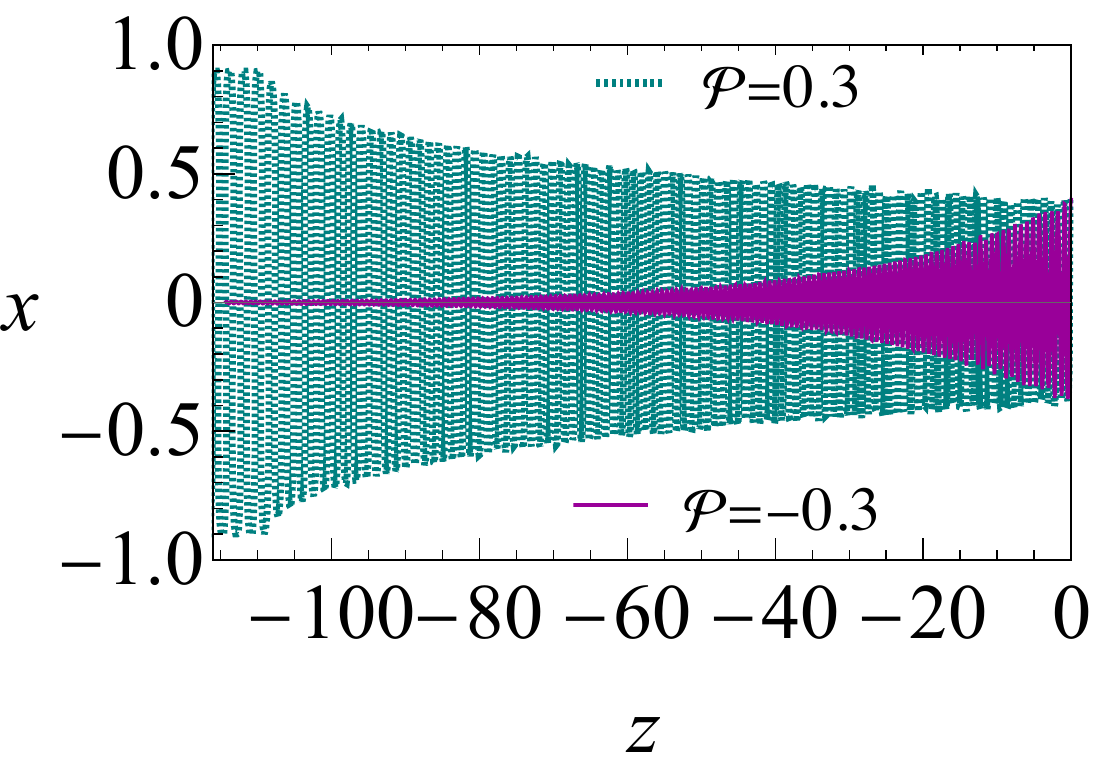} 
		\quad
		\includegraphics[width=0.47\textwidth]{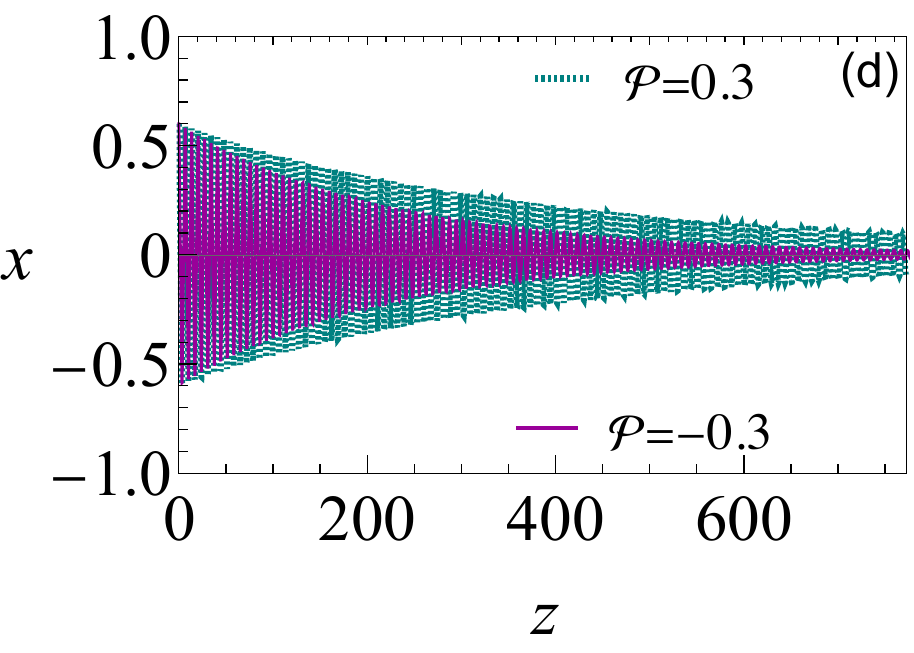}
		\caption{Trajectories showing the centerline focusing of pusher/puller in (a) upstream swimming
			($ \bar{v}_{m}=0.9 $) and (b) downstream 	drifting  ($\bar{v}_{m} = 8 $).  (c) Upstream 
			($ \bar{v}_{m}=0.9 $)
			and (d) downstream trajectories ($ \bar{v}_{m}=3 $)
			with hydrodynamic wall interactions (\ref{wall}) incorporated.
			We employ hard-core  repulsion at the walls.
			Other parameters: $ \text{Wi}=0.1, \, \kappa=0.1, \, \delta=-0.5. $}%
		\label{fig:Traj_FD}%
	\end{figure}
	
	So far we have neglected the hydrodynamic interactions of microswimmers with the bounding channel walls.
	For force-dipole swimmers, the hydrodynamic wall interactions add a modification of order $ \kappa^{2} $ and $ \kappa^{3} $ to the evolution equations of position and orientation, respectively \cite{zottl2012nonlinear,spagnolie_lauga_2012}:
{\small	\begin{align}\label{wall}
		\dot{x} &= -\sin \psi + \mathcal{F} - \frac{3 \mathcal{P} (3 \sin^{2} \psi -1) }{8} \kappa^{2} \left[ \frac{1}{(1-x)^{2}} - \frac{1}{(1+x)^{2}} \right] , \nonumber \\
		\dot{\psi} &= x \bar{v}_{m} - \frac{3 \mathcal{P} \sin 2\psi }{16} \kappa^{3} \left[ \frac{1}{(1-x)^{3}} + \frac{1}{(1+x)^{3}} \right] . 
\end{align}}
	Upstream trajectories in Fig. \ref{fig:Traj_FD}(c) closely resemble the behavior reported previously
	for pure Newtonian fluids \cite{zottl2012nonlinear}. 
	We observe that the hydrodynamic wall attraction of pushers
	\cite{berke2008hydrodynamic}
	overcomes $ \mathcal{F}_{\text{passive}}$ and results in swinging across the whole channel cross section, where the
	strong vorticity near the walls always
	re-orients the swimmer away from it.
	Pullers are repelled from walls \cite{zottl2012nonlinear} and, therefore, rapidly focus  on the centerline.
	In contrast, for downstream drifting
	at large flow rates [Fig. \ref{fig:Traj_FD}(d)], $ \mathcal{F}_{passive} $ dominates 
	over the hydrodynamic wall interactions and all trajectories  tend towards the centerline.
	We note that for source-dipole swimmers
	the hydrodynamic wall interactions are an order $ \kappa $ weaker, and therefore do not alter the trajectories qualitatively.

	
	In conclusion, the current study analyzes microswimmers in weakly viscoelastic pressure-driven flows. 
	For neutral and pusher/puller microswimmers, we derive an additional swimming lift velocity depending on the swimmer's hydrodynamic signature that adds
	to the passive viscoelastic lift \cite{ho1976migration,lee2010cross,mathijssen2016upstream,davino2017particle}. 
	For source-dipole (neutral) swimmers, the swimming lift is two orders of magnitude stronger than the passive lift, which was 
	considered alone in a recent study \cite{mathijssen2016upstream}. The current work shows that 
	the swimming lift accelerates the centerline focusing.
	For force-dipole swimmers (pusher/puller),  the  swimming lift does not contribute to a net cross-streamline migration. 
	Incorporating hydrodynamic wall interactions, we show that upstream swimming for weak flow strengths qualitatively follows the 
	behavior in Newtonian fluids \cite{zottl2012nonlinear}: 
	attraction of pushers towards the channel walls and repulsion of pullers. 
	The downstream drifting along the centerline qualitatively remains the same as that of a passive particle.
	
	These results suggest that normal stresses in viscoelastic fluids generated by the microswimmer's flow field can accelerate the centerline focusing. 
	However, this strongly depends on the hydrodynamic signature of the microswimmer. 
	Even for a weakly viscoelastic fluid ($ \text{Wi} =0.1 $), we observe rapid focusing within 
	a traveled distance of 10-500 times the channel width (Fig. \ref{fig:Traj_SD}), which amounts to ca. $1-50 \text{mm}$ and
	is quite realistic for microfluidic channels.
	Thereby, this work contributes to the understanding of swimming in more realistic biological fluids.
	Furthermore, the current work offers several new directions to explore. For instance, elongated microswimmers perform Jeffery orbits in sheared 
	Newtonian fluids
	\cite{jeffery1922motion}.
	In viscoelastic fluids
	the flow disturbances from swimming will alter the orientation evolution of these orbits
	and  hence the
	swimmer dynamics \cite{corato2017dynamics}.
	Impact of shear-thinning fluids is also an interesting outlook, which can be achieved by the use of more detailed rheological models
	\cite{bird1987dynamics}.\\ \\
	
	Support from Alexander von Humboldt fellowship is gratefully acknowledged.

	\bibliography{Akash,randombib}

	\maketitle
	
	
	\clearpage
	
	\pagebreak

	\onecolumngrid

	\begin{center}
		\textbf{\large Supplemental material for ``On the cross-streamline lift of microswimmers in viscoelastic flows"}
	\end{center}
\bigskip\bigskip

	\twocolumngrid
	
	\setcounter{equation}{0}
	\setcounter{figure}{0}
	\setcounter{table}{0}
	\renewcommand{\theequation}{S\arabic{equation}}
	\renewcommand{\thefigure}{S\arabic{figure}}
	
	

\section{Problem formulation}

\begin{figure}[H]
	\centering
	{{\includegraphics[width=0.43\textwidth]{Schematic} }}%
	\qquad
	{{\includegraphics[width=0.43\textwidth]{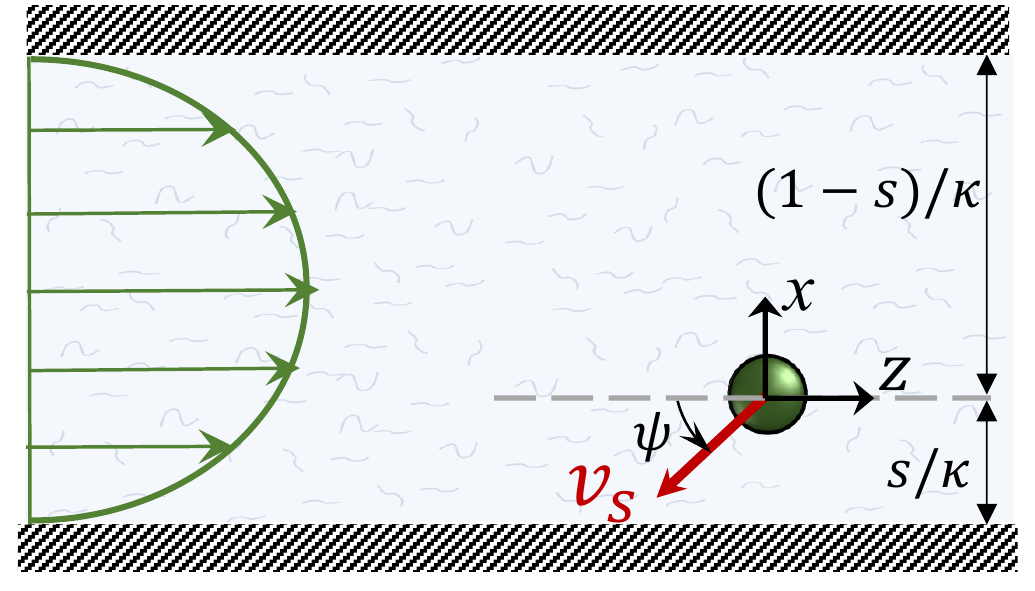} }}%
	\caption{(left) A spherical microswimmer self-propels with velocity $ \IB{v}_{s} = v_s \IB{p}$ in a planar Poiseuille flow inside a channel with half width $w$. The coordinate frame $ \lbrace \tilde{x},\tilde{y},\tilde{z} \rbrace $ co-moves with the swimmer. (right ) Schematic with 
		all lengths normalized by particle radius $ a $. The tilde notation of the coordinates is shown to be dropped for brevity.}
	\label{fig:schematic_supp}%
\end{figure}

The schematic in Fig.\ \ref{fig:schematic_supp} (a) shows a neutrally buoyant spherical microswimmer suspended in pressure-driven flow of a polymeric fluid between two walls. In order to derive the expressions for 
the lift velocities, we work in a reference frame that translates with the swimmer $ {(\tilde{x},\tilde{y},\tilde{z})} $. For simplicity, we temporarily  drop the tilde $ \tilde{} $ notation.
Fig.\ \ref{fig:schematic_supp} (b) shows the non-dimensional description, where $ s  = d/2w $ and $s / \kappa = d /a $
is the distance from the bottom wall normalized by the particle radius $a$.

We split the actual velocity field into the background flow field $\IB{v}^{\infty}$ and the disturbance field $\IB{v}$ (\emph{i.e.} $ \IB{v}_{\text{actual}} = \IB{v} + \IB{v}^{\infty}$), and substitute in Eq.\ (2) of the article.
The inertia-less hydrodynamics of the disturbance field is governed by the continuity and momentum equations in the co-moving swimmer frame $ \lbrace \tilde{x},\tilde{y}, \tilde{z}  \rbrace $ as\footnote{We follow a quasi-steady description because the time scales associated with cross-stream motions both due to swimming	($ a/v_{s} \sim 1\text{s} $) and viscoelastic lift	are much larger than  the characteristic vorticity diffusion time ($ a^{2}/\nu \sim 10^{-4} \text{s}$). }: 
\begin{eqnarray}\label{GE}
	\nabla \cdot \IB{v} &=& 0, \quad  - \nabla p + \nabla^{2} \IB{v} = - \text{Wi} \, (\nabla \cdot \bten{s}), \text{ where} \nonumber \\ 
	\bten{s} &=& 4 (\bten{e} \cdot \bten{e} + \bten{w}) + 2 \, \delta(\overset{\Delta}{\bten{e}}+\overset{\Delta}{\bten{w}}).
\end{eqnarray}
Here, $ \bten{e} $ is the rate of strain tensor for the disturbance flow $ (\nabla \IB{v} + \nabla \IB{v}^{\dagger})/2 $, whereas $ \bten{w} $, $ \overset{\Delta}{\bten{e}} $ and $ \overset{\Delta}{\bten{w}} $ are the different 
parts of the
perturbation $ \bten{s}$ of the polymeric tensor due to the disturbance flow field $\IB{v} $:
\begin{eqnarray}
	\bten{w} &=& \bten{e}^{\infty}\cdot \bten{e} + \bten{e} \cdot \bten{e}^{\infty}, \nonumber \\ 
	\overset{\Delta}{\bten{e}} &=& \IB{v} \cdot \nabla \bten{e} + \bten{e} \cdot \nabla \IB{v}^{\dagger} + \nabla \IB{v} \cdot \bten{e}, \nonumber \\
	\overset{\Delta}{\bten{w}}&=&{\IB{v}^\infty} \cdot \nabla \bten{e} + \bten{e} \cdot \nabla {\IB{v}^\infty}^{\dagger} + \nabla {\IB{v}^\infty} \cdot \bten{e}
	+ \, \IB{v} \cdot \nabla \bten{e}^{\infty} \\ \nonumber
	&&+ \bten{e}^{\infty}\cdot \nabla \IB{v}^{\dagger} + \nabla \IB{v} \cdot \bten{e}^{\infty},
\end{eqnarray}
where $ \bten{e}^{\infty}$  is the rate of strain tensor for the undisturbed flow, $ \overset{\Delta}{\bten{e}} $ is the lower convected 
derivative of $ \bten{e}$ (also known as the Rivlin-Eriksen tensor), $ \bten{w} $ is the `interaction tensor' (arising from the interaction between background flow and disturbance field), and $ \overset{\Delta}{\bten{w}} $ is 
its lower convected derivative.

The above equations are non-dimensionalized using $ a , \, \kappa v_{m} , \, \mu  \kappa v_{m} /a  $ as the characteristic scales for length, velocity, and pressure, respectively. The definitions of these dimensional parameters $ a $ (particle size), $ \kappa = a/2w $,
and $ v_{m} $ (maximum flow velocity) are consistent with the communication article.
In our case, $ \IB{v}^{\infty} $ is the undisturbed Poiseuille flow velocity in the frame of reference translating with the particle
\begin{equation}
	{\IB{v}^\infty } = \left( {\alpha  + \beta x + \gamma {x^2}} \right){\IB{e}_z}  - {\IB{U}_p},
	\label{A-VInf}
\end{equation}
where $ \IB{U}_{p} $ is the total velocity of the swimmer, \emph{i.e.}, swimming velocity $\IB{v}_s$ plus advection due to the Poiseuille flow and the lift velocities.
The constants $ \alpha, \beta$ and $\gamma $ are:
\begin{equation}
	\alpha  = 4 s\left( {1 - s} \right)/\kappa ,  \, \beta  = 4 \left( {1 - 2s} \right), \,\gamma  =  - 4{\kappa },
	\label{alpha}
\end{equation}
where $\beta$ and $\gamma$ represent the shear and  curvature of the background flow, respectively. 
The boundary conditions of the disturbance flow field are
\begin{subequations}\label{BC}
	\begin{gather}
		\IB{v} =  \IB{v}_{\theta} + {\IB{\Omega}_s} \times \IB{r} - \IB{v}^{\infty} \quad \mbox{at\ } r=1,\\
		\IB{v} = 0   \quad \mbox{at walls,\ }\\
		\IB{v} \rightarrow \IB{0} \quad \mbox{as\ } \lbrace y,z \rbrace\rightarrow \infty.
	\end{gather}
\end{subequations}
Here, the walls are located at $ x=-s/\kappa $ and $ x=(1-s)/\kappa $, and $ \IB{v}_{\theta} $ represents the prescribed
tangential surface velocity of the 
spherical microswimmer.

\section{Perturbation expansion}
We find the viscoelastic lift or migration velocities at $ O(\text{Wi}) $ using a regular perturbation expansion. For small values of $ \text{Wi} $, the disturbance field variables are expanded as:
\begin{equation}
	\xi=\xi_{0} + \text{Wi} \; \xi_{1} + \cdots.
	\label{pert}
\end{equation}
Here, $ \xi $ is a generic field variable which represents velocity ($ \IB{v} $), pressure ($ p $), translational ($ \IB{U}_{p} $) and angular velocity ($ \IB{\Omega}_{p} $). 
We substitute (\ref{pert}) in the equations governing the 
disturbance field (\ref{GE}) and obtain the problem at $ O(1) $ (\emph{i.e.} Stokes problem) as
\begin{equation}
	\left. \begin{array}{l}
		\quad  \; \; \, \quad \nabla  \cdot {\IB{v}_{0}} = 0, \\
		{\nabla ^2}{\IB{v}_{0}} - \nabla {p_{0}} = \IB{0},\\
		\qquad \quad \; \quad {\IB{v}_{0}} =  \IB{v}_{\theta} +  
		{\IB{\Omega} _{p0}}    
		\times \IB{r} - {\IB{v}_0^\infty} \quad \mbox{at\ } r = 1,\\
		\qquad \quad \; \quad{\IB{v}_{0}} = 0 \quad \mbox{at walls},\\
		\qquad \quad \; \quad{\IB{v}_{0}} \to \IB{0} \quad \mbox{as\ } \lbrace y,z \rbrace\rightarrow \infty.
	\end{array} \right\}
	\label{Order0}
\end{equation}
and at $ O(\text{Wi}) $ as:
\begin{equation}
	\left. \begin{array}{l}
		\qquad \; \quad \nabla  \cdot {\IB{v}_{1}} = 0, \\
		\; \,\,	{\nabla ^2}{\IB{v}_{1}} - \nabla {p_{1}} =   - \nabla \cdot \bten{s}_{0}   ,\\
		\qquad \qquad \quad {\IB{v}_{1}} = 
		{\IB{U}_{p1}}       
		+      {\IB{\Omega}_{p1}}     
		\times \IB{r}\quad \mbox{at\ } r = 1,\\
		\qquad \qquad \quad{\IB{v}_{1}} = \IB{0} \quad \mbox{at walls},\\
		\qquad \qquad \quad{\IB{v}_{1}} \to \IB{0} \quad \mbox{as\ }  \lbrace y,z \rbrace\rightarrow \infty. 
	\end{array} \right\}
	\label{Order1}
\end{equation}
In (\ref{Order0}), $ \IB{v}^{\infty}_{0} = \left( {\alpha  + \beta x + \gamma {x^2}} \right){\IB{e}_z}  - {\IB{U}_{p\,0}} $.

\citet{ho1976migration}, in their seminal work, used the reciprocal theorem to derive a volume integral expression for the
migration velocity associated with the $ O(\text{Wi}) $ equations (\ref{Order1}):
\begin{equation}
	U_{\text{mig}} \equiv \text{Wi} \, \IB{U}_{p1} \cdot \IB{e}_{x} =  - \frac{1}{6\pi} \text{Wi} \; \int_{V_{f}}  \bten{s}_{0} : \tilde{\nabla} 
	\IB{v}^{t} \: \rm{d}V.
	\label{mig}
\end{equation}
The auxiliary or test field ($ \IB{v}^{t},\,p^{t} $) is associated with a sphere moving in the positive $x$-direction (towards the upper wall) with unit velocity in a quiescent fluid: 
\begin{equation}
	\IB{v}^t(\IB{r})  
	= \frac{3}{4}\left( {{\IB{e}_x} + \frac{{x\IB{r}}}{{{r^2}}}} \right)\frac{1}{r} + \frac{1}{4}\left( {{\IB{e}_x} - \frac{{3x\IB{r}}}{{{r^2}}}} \right)\frac{1}{{{r^3}}}.
	\label{Test_Lamb}
\end{equation}
The reciprocal theorem makes it relatively easy to find 
lift velocities at $O(\text{Wi})$,
as we can solve the creeping flow problem (\ref{Order0}) using well-established methods \cite{lamb,kim2013} and directly substitute its solution in (\ref{mig}). 
In other words, we do not need to
solve the $ O(\text{Wi}) $ problem (\ref{Order1}) to obtain the $ O(\text{Wi}) $ lift.

\section{Viscoelastic lift velocity: Source-dipole swimmer}

We now use 
expression (\ref{mig}) for evaluating the swimming lift of  a source-dipole swimmer. We explicitly choose the axisymmetric neutral squirmer, which has the surface velocity field 
$ \IB{v}_{\theta} = B_1 \sin \theta \IB{e}_\theta$, where $\theta$ is the polar angle and $\IB{e}_\theta$ the corresponding base vector.
The swimming velocity is directly related to this squirmer coefficient: $v_s =  2B_1/3
$ \cite{ishikawa2006hydrodynamic,zottl2016emergent}.
The solution to the $ O(1) $ Stokes problem (\ref{Order0}) can be divided in swimming and passive disturbances.
From the squirmer model \cite{blake1971spherical}, we obtain:
\begin{eqnarray}
	\IB{v}^{(1)\text{swim}}_{0} &=& \frac{v_s \IB{p} }{2 r^{3}} 	
	\IB{\cdot} \left[    \frac{3 \IB{r} \IB{r} }{r^{2}}   -\bten{I} \right]   	\nonumber\\
	= && \frac{v_{s}}{2r^{3}} \left[  \cos \psi \left( \frac{3 z \IB{r}}{r^{2}} - \IB{e}_{z}  \right) 
	\; +\;    \sin \psi \left( \frac{3 x \IB{r}}{r^{2}} - \IB{e}_{x}   \right) \right].\nonumber \\ \label{dist_neut}
\end{eqnarray}
Using Lamb's general solution \cite{lamb}, we obtain the passive disturbance
\begin{widetext}
\begin{eqnarray}
	\IB{v}^{\text{passive}}_{0} &=&  {B}\left( { - {\IB{e}_z} + \frac{{3z\IB{r}}}{{{r^2}}}} \right)\frac{1}{{{r^3}}} + {D}{\frac{{zx\IB{r}}}{{{r^5}}}} +  E \left(x \IB{e}_{z} + z \IB{e}_{x} - \frac{5 x z \IB{r}}{r^{2}}\right) \frac{1}{r^{5}}
	\nonumber\\
	&+& F \left(\IB{e}_{z} - \frac{2x^{2}\IB{e}_{z} + z\IB{r}}{r^{2}} + \frac{2xz\IB{e}_{x}}{r^{2}} \right)\frac{1}{r^{3}}  + G\left(\IB{e}_{z} - \frac{5x^{2}\IB{e}_{z} +10xz\IB{e}_{x} + 13 z\IB{r} }{r^{2}} + \frac{75zx^{2} \IB{r}}{r^{4}} \right)\frac{1}{r^{3}}   \nonumber\\
	&+&  H\left(\IB{e}_{z} - \frac{5x^{2}\IB{e}_{z} +10xz\IB{e}_{x} + 5 z\IB{r} }{r^{2}} + \frac{35zx^{2}\IB{r}}{r^{4}} \right)\frac{1}{r^{5}} \, , \label{Lamb_vel}
\end{eqnarray}
\end{widetext}
where the coefficients are defined as:
{\small \begin{eqnarray}
	&&  \; B= \frac{\gamma}{15} , \;
	D=-\frac{5\beta}{2},  \;  E=-\frac{\beta}{2} ,\; F=\frac{\gamma}{3}, \;  G= -\frac{7 \gamma}{120} ,\; H=\frac{\gamma}{8}  . \nonumber \\
	\label{coeff}
\end{eqnarray}}
The terms multiplying the coefficients  $ B $, $ D $, $ E $ represent 
source-dipole,  stresslet, and octupole singularities, respectively  \citep{kim2013,guazzelli2011}. 
The other disturbances (terms multiplying $ F,\, G,\, H $) are further singularities in the multipole expansion, which arise due to the  curvature $ \gamma $
in the background flow field together with the source dipole.

The tensor $\bten{e}^{\infty} $ is yet unknown for the Poiseuille flow  of Eq.\ (\ref{A-VInf}) in zeroth order of $\text{Wi}$.
To calculate it, we note that the total velocity of the force-free swimmer in the Stokes regime is $ \IB{U}_{p\,0} = \IB{v}_{s} + (\alpha + \gamma/3  )\IB{e}_{z} $ (the second part is obtained by using Faxen's laws \cite{kim2013}). To complete the expression of $ {\IB{v}_{0}^{\infty}} $, we substitute $ \IB{U}_{p\,0} $ in (\ref{A-VInf}), and obtain:
\begin{equation}\label{v_inf}
	{\IB{v}_{0}^{\infty}}  = (\beta x+ \gamma x^{2} - \gamma/3 )\IB{e}_{z} -\IB{v}_{s}
\end{equation}
which gives  $[\bten{e}^{\infty}]_{xz} = [\bten{e}^{\infty}]_{zx} = (\beta + 2 x \gamma)/2 $.

Now, we evaluate the 
volume integral  in Eq.\ (\ref{mig}).
Since  the source-dipole field of the neutral swimmer decays quickly away from the swimmer ($ \sim 1/r^{3} $), we can neglect the wall 
corrections in the 
volume integral of Eq.\ (\ref{mig}).
In the context of  an electrophoretic source-dipole disturbance, \citet{choudhary2020electrokinetically} showed that the error generated from this neglection is dispensable.
Integrating over the infinite space, we obtain the swimming lift velocity in units of $v_s$ as
\begin{equation}\label{SD_lift}
	U_{\text{mig}} =  \text{Wi}  \, \left[ \, (5/9) \beta \gamma (1+3\delta) \bar{v}_{m} \kappa \, + \,  (1/4) \beta (1+\delta) \cos \psi \, \right] ,
\end{equation}
expressed in the co-moving frame of the swimmer.
The first component is the passive lift velocity (identical to that obtained by \citet{ho1976migration}), and 
the second component is the swimming-lift velocity that arises due to the source-dipole disturbance created by the neutral swimmer. 

\section{Viscoelastic lift velocity: Force-dipole swimmer}

As before, $ \IB{v}_{0} $ is the combination of flow fields due to swimming and the passive disturbance. 
The latter is identical to (\ref{Lamb_vel}); for the 
swimmer, we take the force-dipole field from the studies on flagellated microswimmers \cite{lauga2004,spagnolie_lauga_2012}, where 
$ \mathcal{P} $ is the dipole strength normalized with  $ 8\pi \mu a^{2} v_{s} $:
\begin{widetext}
\begin{eqnarray}
	\IB{v}^{(1)swim}_{0} &=&  \mathcal{P}  \IB{r} \left[   \frac{-1}{r^{3}} + 3 \frac{\left(\IB{r} \cdot \IB{p}  \right)^{2} }{r^{5 }}  \right]
	\nonumber\\
	&=&   \mathcal{P} \cos^{2} \psi \left( \frac{-\IB{r}}{r^{3}} + \frac{3z^{2} \IB{r}}{r^{5}}  \right) 
	\; +\;   \mathcal{P}  \sin^{2} \psi \left( \frac{-\IB{r}}{r^{3}} + \frac{3x^{2} \IB{r}}{r^{5}}  \right) 
	\; +\;   \mathcal{P}  \sin 2\psi \left( \frac{3x  z \, \IB{r}}{r^{5}} \right). \label{Lamb_vel_active}
\end{eqnarray}
\end{widetext}
Substituting the above equation 
together with Eq. (\ref{Lamb_vel})  into Eq.\ (\ref{mig})
and integrating over the infinite domain, we obtain (in the units of $ v_{s} $):

{\small \begin{equation}\label{FD_lift}
	U_{\text{mig}} =  \text{Wi}  \, \left[ \, (5/9) \beta \gamma (1+3\delta) \bar{v}_{m} \kappa  \, - \,  (2/3)  \mathcal{P}   \gamma (1+3\delta) \sin 2\psi \, \right].
\end{equation}}
The second component is the additional swimming-lift velocity that will be experienced by the force-dipole swimmer.
Note that it depends on the curvature $\gamma$ of the Poiseuille flow.

\section{Inertial lift velocities in the channel frame}
\label{sect.inertial_channel}

Here we provide the final expressions of  the swimming and passive lift  velocities in the channel frame of reference, which is
used in the communication article. The conversion requires a transformation of particle-wall distance $ s / \kappa $ to the channel $x$ coordinate (see Fig.\ \ref{fig:schematic});
for $x$ in units of $w$ we then have $s = (1+x)/2$.
\medskip

\noindent
\textbf{1. Swimming lift: Neutral microswimmer}\\
Using the definition of $ \beta $ (\ref{alpha}) and 
$s = (1+x)/2$
in (\ref{SD_lift}), yields:
\begin{equation}
	U_{\text{mig}} = \mathcal{F} _{\text{swim}} = - \text{Wi} (1+\delta) \, x  \cos \psi .
\end{equation}

\medskip

\noindent
\textbf{2. Swimming lift: Pusher/puller microswimmer}\\
Using the definition of $ \gamma $ (\ref{alpha}) 
in  (\ref{FD_lift}), yields:
\begin{equation}
	U_{\text{mig}} = \mathcal{F} _{\text{swim}} =\text{Wi}  (8/3) (1+3\delta) \,  \mathcal{P} \kappa\sin 2 \psi
\end{equation}

\noindent
\textbf{3. Passive lift}:\\
Using the definition of $ \beta, \gamma $ (\ref{alpha}) and  $s = (1+x)/2$ in the passive lift component, yields:
\begin{equation}\label{passive_fit}
	U_{\text{mig}}  = \mathcal{F}_{\text{passive}} = \text{Wi}  (80/9) (1+3\delta) x \, \bar{v}_{m} \kappa^{2}  .
\end{equation}

\section{Particle drift and rotation modification}
\label{sect.extra}

To calculate the viscoelastic modification to drift and rotational velocity of the swimmer, we use the following two test fields (respectively):
\begin{equation}\label{drift_test}
	\IB{v}^t(\IB{r})  
	= \frac{3}{4}\left( {{\IB{e}_z} + \frac{{z\IB{r}}}{{{r^2}}}} \right)\frac{1}{r} + \frac{1}{4}\left( {{\IB{e}_z} - \frac{{3z\IB{r}}}{{{r^2}}}} \right)\frac{1}{{{r^3}}}.
\end{equation}

\begin{equation}\label{rot_test}
	\IB{v}^t(\IB{r})  
	= \frac{x \IB{e}_{z}  -  z \IB{e}_{x}}{r^3}.
\end{equation}

\subsection{Source-dipole swimmer}
Substituting (\ref{drift_test}) in the volume integral (\ref{mig}), we obtain the drift modification as:
\begin{equation}
	U_{drift}= \frac{1}{4} \text{Wi} \, \beta \, v_{s} (1+\delta) \sin \psi.
\end{equation}
Substituting (\ref{rot_test}) in the volume integral (\ref{mig}), we obtain the rotation modification as:
\begin{equation}
	\IB{\Omega}_1 = \frac{2}{3} \text{Wi} \, \gamma \, v_{s} \sin \psi \, \IB{e}_{y} 
\end{equation}

\subsection{Force-dipole swimmer}
Similarly, for force-dipole swimmer, we obtain 
\begin{equation}
	U_{drift}= \frac{1}{12} \text{Wi} \, \gamma \, v_{s} (1+\delta) (-3+5\cos \psi), \nonumber 
\end{equation}
\begin{equation}
	\IB{\Omega}_1 = \frac{2}{3} \text{Wi} \, \beta \, v_{s} (1+3 \delta) \cos 2\psi.\nonumber 
\end{equation}
Since these effects are an order of magnitude (in $ \text{Wi} $) smaller than the flow speed and flow vorticity (respectively), they do not alter the swimmer dynamics.


\end{document}